\documentclass[%
 pre,reprint,
 superscriptaddress,
 amsmath,amssymb,
 aps,
]{revtex4-2}

\usepackage{graphicx}
\usepackage{bm}
\usepackage{latexsym}
\usepackage{xcolor}
\usepackage{svg}
\usepackage{cancel}
\usepackage{enumitem}
\usepackage[hidelinks]{hyperref}
\usepackage{tikz}
\usepackage{color}

\begin{document}

\title{Reciprocal swimming in viscoelastic granular hydrogels}

\author{Hongyi Xiao$^{\dag}$}
\affiliation{Institute of Multiscale Simulation, Friedrich-Alexander-Universität Erlangen-Nürnberg, Erlangen, Germany.}
\affiliation{Department of Mechanical Engineering, University of Michigan, Ann Arbor, USA}%

\author{Jing Wang$^\dag$}
\affiliation{Institute of Physics, Otto von Guericke University Magdeburg, Magdeburg, Germany.}
\thanks{These two authors contributed equally}

\author{Achim Sack}
\affiliation{Institute of Multiscale Simulation, Friedrich-Alexander-Universität Erlangen-Nürnberg, Erlangen, Germany.}
\author{Ralf Stannarius}
\affiliation{Institute of Physics, Otto von Guericke University Magdeburg, Magdeburg, Germany.}
\affiliation{Brandenburg University of Applied Sciences, Brandenburg a.\,d.\,Havel, Germany.}

\author{Thorsten Pöschel}
\email{thorsten.poeschel@fau.de}
\affiliation{Institute of Multiscale Simulation, Friedrich-Alexander-Universität Erlangen-Nürnberg, Erlangen, Germany.}

\begin{abstract}
We experimentally study a scallop-like swimmer with reciprocally flapping wings in a nearly frictionless, cohesive granular medium consisting of hydrogel spheres. Significant locomotion is found when the swimmer's flapping frequency matches the inverse relaxation time of the material. Remarkably, the swimmer moves in the opposite direction compared to its motion in a cohesion-free granular material of hard plastic spheres. 
At higher or lower frequencies, we observe no motion of the swimmer, apart from a short initial transient phase. X-ray radiograms reveal that the wing motions create low-density zones, which in turn give rise to a hysteresis in drag and propulsion forces. This time-dependent effect, combined with the swimmer’s inertia, accounts for locomotion at intermediate frequencies.
\end{abstract}

\maketitle

\section{Introduction}
\label{sec:intro}

Understanding active locomotion, such as the propulsion of living matter \citep{goldstein2016batchelor, stephen2012natural, vogel2020life} or active colloids \citep{buttinoni2013dynamical, herminghaus2014interfacial, ebbens2014electrokinetic}, is a challenging problem. In a Newtonian liquid at low Reynolds numbers, effective locomotion requires an asymmetrical sequence of shape changes. Reciprocal deformation cannot cause net propulsion due to the linearity and time reversibility of the Stokes equation, explained by Purcell's \textit{scallop theorem} \citep{purcell1977life}. In non-Newtonian liquids with nonlinear rheological properties such as shear-dependent viscosity and normal-stress differences, swimmers are not constrained by this theorem, as they can achieve locomotion by periodic asymmetrical and symmetrical shape changes, e.g., cells with waving flagella and flapping scallops \citep{lauga2011life, arratia2022life, qiu2014swimming, xu20243d}. 

Non-continuum effects play an important role in complex fluids with heterogeneous microstructures, such as colloids, polymer solutions, and granular materials \citep{balin2017biopolymer, zhang2018reduced}. 

In highly concentrated polymer solutions, molecules can form gel-like networks, resulting in nonlinear rheological properties of the fluid. Swimmers can take advantage of effects such as normal stresses that increase with shear rate \citep{normand2008flapping, pak2012micropropulsion, datta2022perspectives} and shear-rate–dependent viscosity arising from body movements \citep{fu2009swimming, qiu2014swimming, riley2017empirical}. Propulsion gaits that stretch polymers in shear flow, thereby reducing the local viscosity around the swimmer, can enhance swimming efficiency under these conditions \citep{fu2010low, wrobel2016enhanced}. In these situations, the flow of polymeric liquids can still be understood within the framework of fluid mechanics.

Locomotion in granular media is a particularly rich problem, with examples ranging from sandfish lizards that undulate through sand \citep{maladen2009undulatory, goldman2014colloquium}, to worms that burrow in soil via peristalsis \citep{dorgan2018kinematics}, and southern sand octopuses that fluidize the substrate to alter the resistance of submerged sand \citep{montana2015liquid}. These systems are challenging to understand because granular matter can behave either like a flowing liquid \citep{forterre2008flows, xiao2017transient, xiao2018unsteady} or like a load-bearing solid \citep{majmudar2007jamming, bi2011jamming}, depending on the conditions. The solid-like behaviour arises from the jamming transition, which can be induced by isotropic pressure or shear \citep{liu1998jamming, bi2011jamming, behringer2018physics}, and has important implications for locomotion. For instance, particles ahead of an intruder in a static bed often remain jammed, effectively enlarging the penetrating object \citep{Mueller:2014, Buchholtz:1998, brzinski2013depth, harrington2020stagnant, kang2018archimedes}. At large stresses, collective particle rearrangements dissipate elastic energy, giving rise to elasto-plastic behaviours \citep{schall2010shear, xiao2020strain, xiao2023identifying}. Most studies on locomotion, therefore, model granular materials as elasto-plastic or perfectly plastic continua, for which reduced-order approaches such as resistive force theory are applicable \citep{askari2016intrusion, agarwal2021surprising, agarwal2023mechanistic}. 

Granular materials containing highly deformable, cohesive, or nearly frictionless particles can display pronounced viscoelasticity, as in food products, powders, cosmetics, and crude oil. Their flow behaviour differs markedly from that of model materials, such as dry glass or plastic beads and poppy seeds, which are typically used in laboratory studies. Granular hydrogels provide a particularly interesting cohesive model system because they combine very low friction coefficients ($\approx 0.01$) with high elastic deformability (elastic moduli $0.01\dots 1\,\text{kPa}$) \citep{Ashour2017, shakya2024viscoelastic}. Due to their high water content, liquid capillary bridges can form between hydrogel particles, inducing cohesive forces. These effects lead to distinct time-dependent behaviours in shear flows \citep{wang2022characterization, Dijksman2022, Farmani2025} and silo discharge \citep{harth2020intermittent, wang2023force, Stannarius2019}. Because granular hydrogels combine viscoelasticity, reminiscent of polymeric liquids, with reconfigurable heterogeneous packing, swimming in such media is of particular interest. Beyond fundamental physics, it may also be relevant for applications such as microrobots navigating among blood cells.

This study presents experiments with a scallop-inspired swimmer driven by two counter-rotating blades that flap reciprocally. The swimmer moves within a bed of hydrogel spheres and, as will be shown, generates propulsion only within a specific frequency range of flapping. This behaviour contrasts to swimmers in beds of frictional polystyrene spheres \citep{Hongyi:2024}, where both the swimming direction and its frequency dependence differ from those observed in rigid frictional particles.

\section{Experimental methods and materials}
\label{sec:exp}

\begin{figure}[htbp]
\centerline{\includegraphics[width=0.95\linewidth]{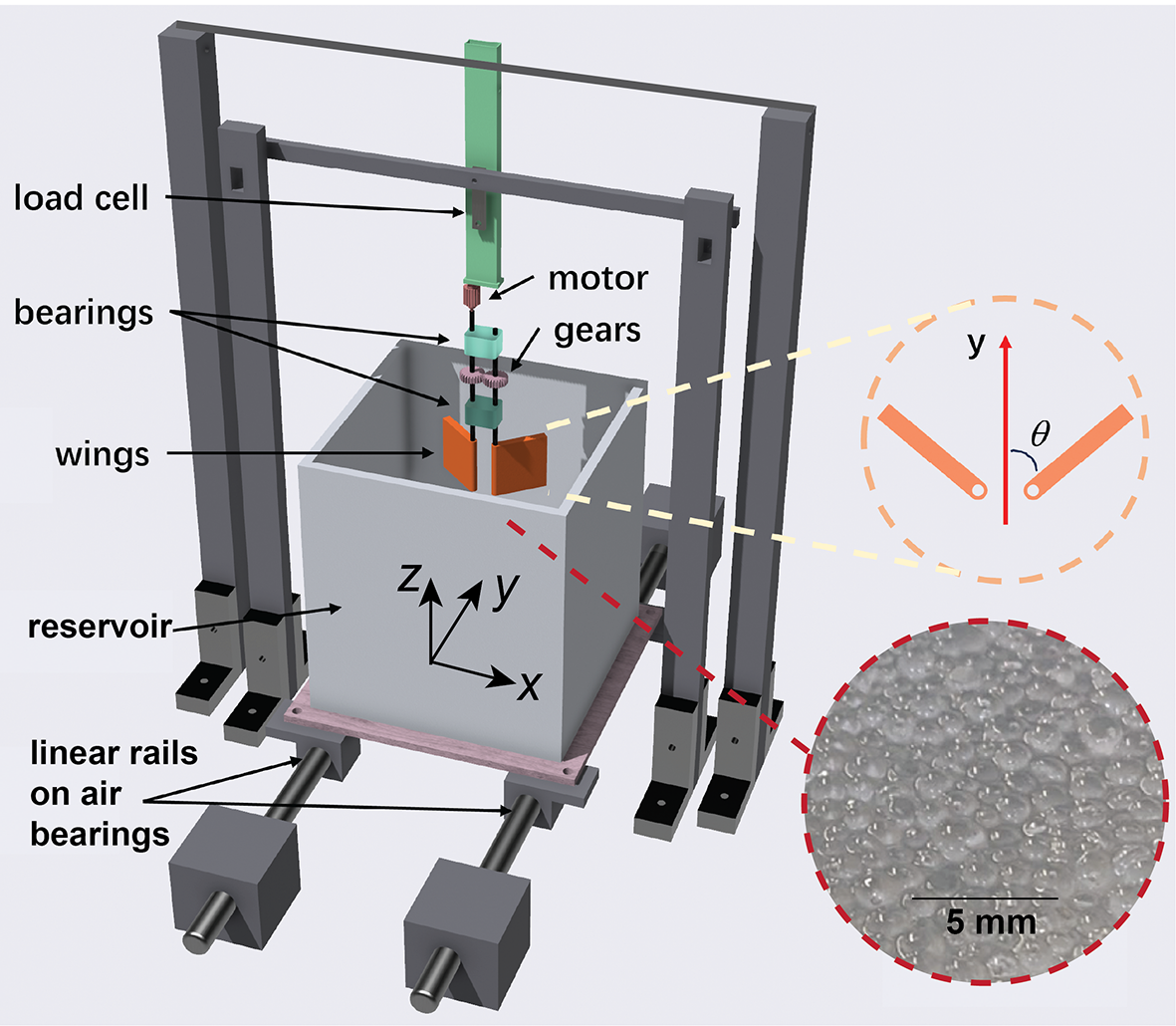}}
\caption{\label{fig:exp} Experimental setup for swimming in granular media. The apparatus consists of a pair of wings operating in a reservoir filled with granular particles. The inset shows a photograph of the hydrogel particles. 
}
\end{figure}

The experimental setup for the scallop-inspired swimmer in granular media is shown in \autoref{fig:exp} based on our previous study \citep{Hongyi:2024}. 
The swimmer consists of a pair of counter-rotating, square wings ($15\,\text{mm}\times\text{15}\,\text{mm}$, thickness $1\,\text{mm}$), 3D printed from resin. Each wing is mounted on a vertical carbon-fibre rod (diameter $2\,\text{mm}$), supported by ball bearings, with the rods separated by $10\,\text{mm}$. A servo motor drives one rod directly, while the other is coupled through a pair of identical gears, ensuring symmetric counter-rotation of the wings. A Cartesian coordinate system is defined with $x$ the lateral direction, $y$ the swimming direction (\autoref{fig:exp} inset), and $z$ the vertical axis pointing upward. The wings flap between  $\theta_1=80^\circ$ (opened) and $\theta_2=20^\circ$ (closed), respectively.

To prepare the hydrogel granulate, we started from dry powder sieved to a narrow size distribution $[0.20, 0.23]\,\text{mm}$. After swelling in distilled water, the particles assumed diameters $[1,2]\,\text{mm}$. The hydrogel particles were removed from the water bath with a strainer and placed on paper towels to drain excess surface water. Residual water formed liquid bridges between neighbouring particles, rendering the material slightly cohesive. The inset of \autoref{fig:exp} shows the bed surface. For comparison, we also used dry polystyrene particles with an average diameter of $1\,\text{mm}$ and the same density as the hydrogel spheres, as examined in greater detail in \citep{Hongyi:2024}.

The granulate was contained in a cubic box with a cross-section of $70\,\text{mm} \times 70~\text{mm}$, filled to a height of $85\,\text{mm}$. The swimmer's mid-$xy$ plane was thus positioned at a depth of $40\,\text{mm}$. The box rested on two linear rails supported by air bearings, allowing frictionless motion in the $y$ direction. The total moving mass, including the reservoir and rails, was $1.54\,\text{kg}$. 

The displacement of the reservoir was recorded by a linear position Hall encoder and converted to the relative displacement of the swimmer, $y_s$. A load cell measured the force $f_y$ acting on the swimmer in the $y$ direction. Neither the load cell nor the motor was mechanically coupled to the moving granular bed. Swimming was tested in both hydrogel and polystyrene media over a range of cycle periods $T$, always beginning from the closed angle $\theta_2$.

\section{Locomotion in granular media of hydrogel and polystyrene particles} \label{free}
\label{sec:loc}

\begin{figure}[htbp]
\centerline{\includegraphics[width=0.95\linewidth]{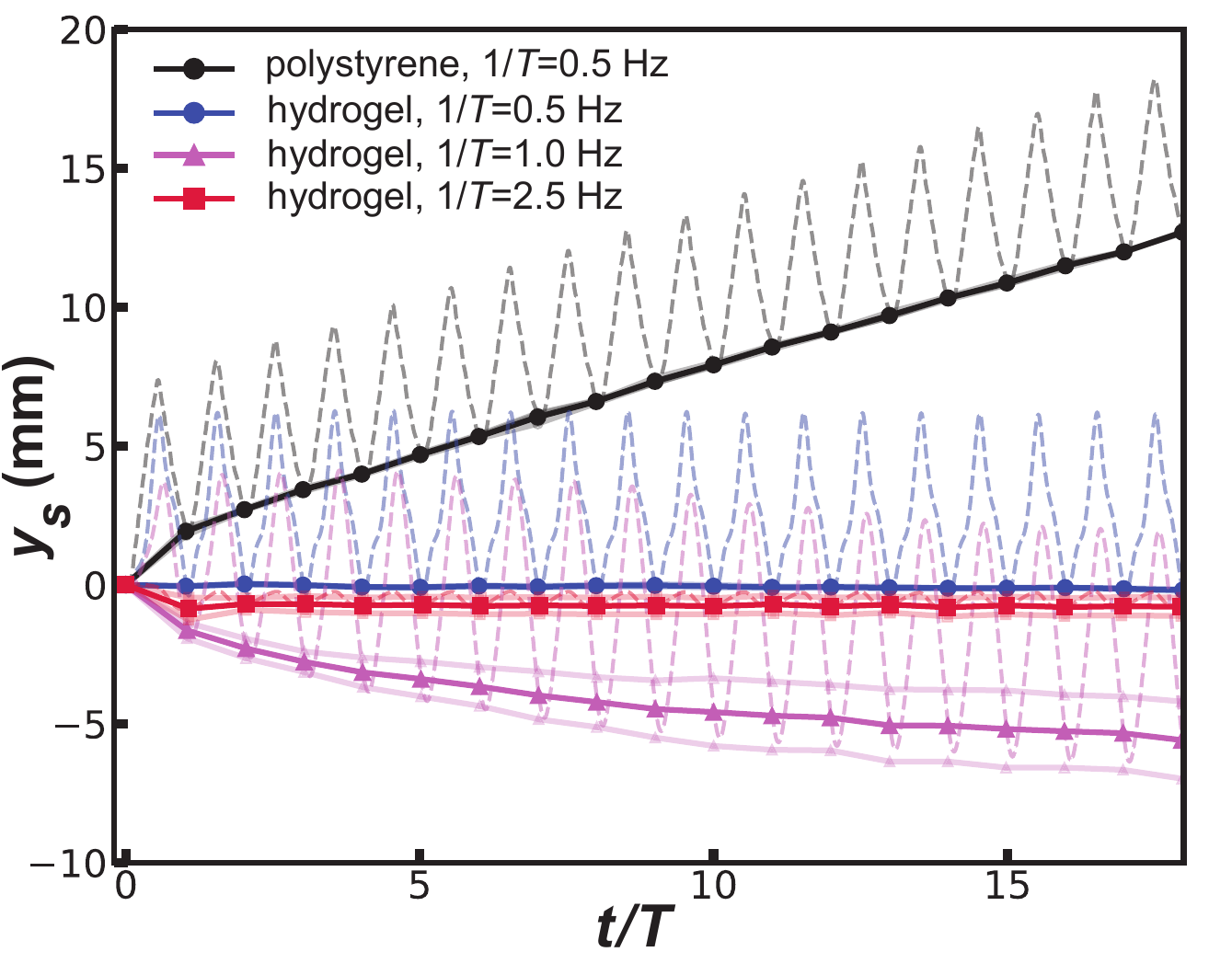}}
\caption{\label{fig:dispA} 
Swimmer displacement as a function of time (dashed lines) for several flapping periods: $T=2.0\,\text{s}$ (blue), $T=1.0\,\text{s}$ (purple), $T=0.4\,\text{s}$ (red). For comparison, the black line shows data for polystyrene granulate at $T=2.0\,\text{s}$. Symbols and solid lines indicate the displacement at the end of each cycle, averaged over two repetitions. }
\end{figure}

\autoref{fig:dispA} shows the swimmer displacement $y_s$ as a function of scaled time $t/T$ (dashed curves), with the positive direction (forward) defined in \autoref{fig:exp}.
In polystyrene granulate, the swimmer exhibits persistent locomotion in the $+y$ direction, consistent with our previous results \citep{Hongyi:2024}. In contrast, locomotion in hydrogel granulate occurs in the $-y$ direction and depends strongly on the flapping rate: at $1/T=0.5\,\text{Hz}$, the net displacement vanishes, with the swimmer returning to $y_s=0$ after each cycle; at $1/T=1.0\,\text{Hz}$ the swimmer develops persistent locomotion in the $-y$ direction, opposite to that observed in polystyrene granulate; and at $1/T=2.5\,\text{Hz}$, the net displacement again vanishes.

\begin{figure}[htbp]
\centerline{\includegraphics[width=0.95\linewidth]{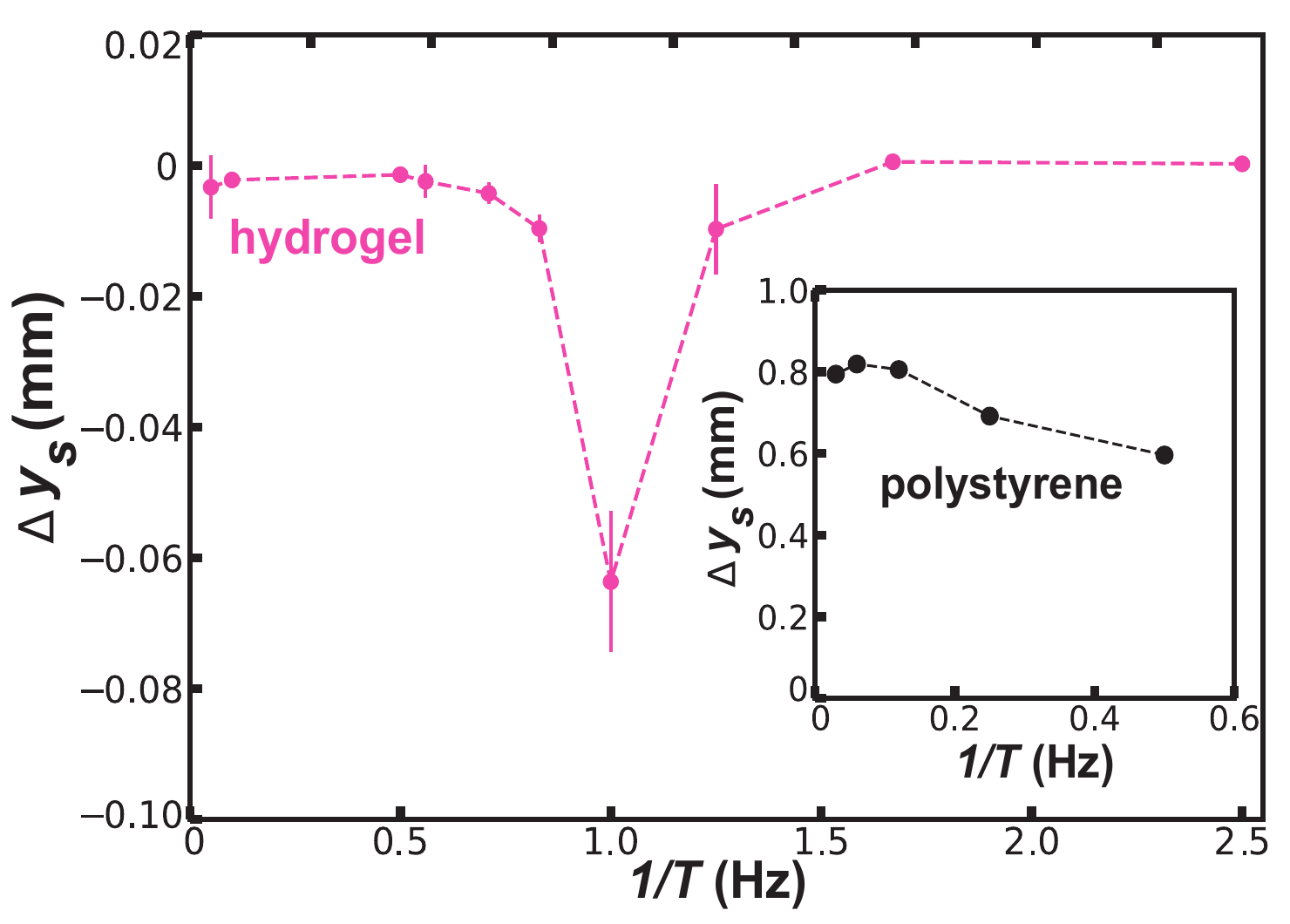}}
\caption{\label{fig:dispB}  
Net swimmer displacement per cycle as a function of the flapping frequency. The inset shows the data for a swimmer moving in a polystyrene granulate; here, the error bars are very small.}
\end{figure}

\autoref{fig:dispB} shows the swimmer's net displacement per cycle, $\Delta y_s$, as a function of the flapping frequency.
These values are averages of the slope of $y_s(t/T)$ at the end of each cycle (symbols in \autoref{fig:dispA}), quantified by linear fits over at least five successive data points in the linear regime after the initial relaxation ($t/T\ge 5$). The error bars are the standard deviations of these data. We obtain a sharp peak at $1/T=1\,\text{Hz}$, while for both $1/T\ll 1\,\text{Hz}$ and $1/T\gg 1\,\text{Hz}$, the motion ceases. For comparison, the inset of \autoref{fig:dispB} shows the same data for a swimmer moving in a polystyrene granulate. Although we cannot scan the same frequency range as for hydrogel particles due to limitations of the apparatus, we see that the net displacement per cycle is strictly positive and decreases slightly with increasing frequency $1/T$. This behaviour resembles the rate-independence reported for other locomotion problems in plastically deforming granular media \citep{maladen2009undulatory, hatton2013geometric}. Whereas the persistent forward motion in stiff, frictional, and non-cohesive granular media is associated with particle jamming in the vicinity of the swimmer \citep{Hongyi:2024}, the backward propulsion observed in hydrogel particles arises from a different mechanism, as discussed below.

\begin{figure}[htbp]
\centerline{\includegraphics[width=0.95\linewidth]{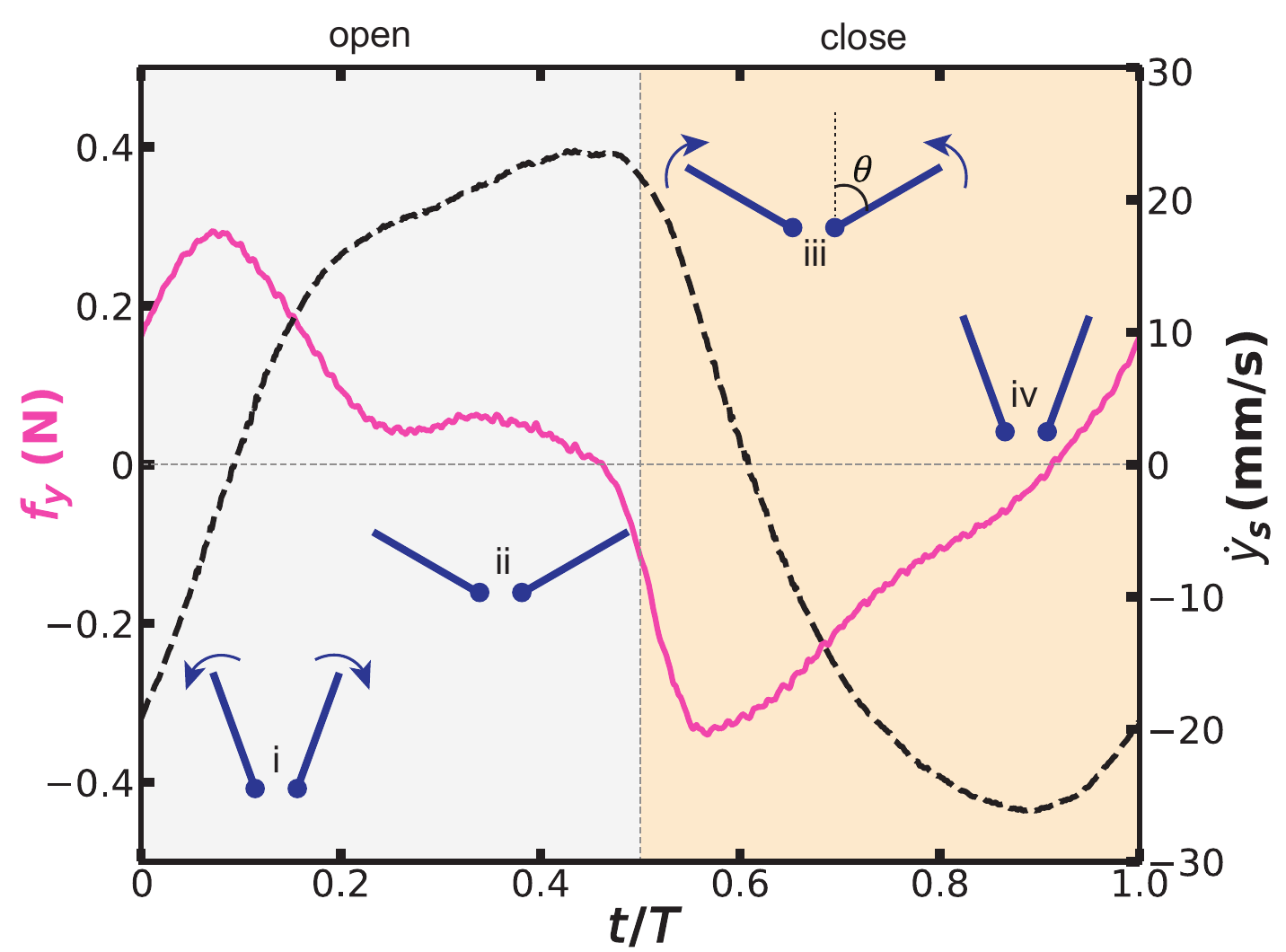}}
\caption{\label{fig:dispC} 
Phase-resolved swimmer velocity (black dashed line, right axis) and driving force (red solid line, left axis) at a flapping frequency of $1/T=1\,\text{Hz}$. The vertical dotted line separates the opening and closing half-cycles, sketched in the insets.}
\end{figure}

While \autoref{fig:dispB} presents the average swimmer's net displacement per cycle, \autoref{fig:dispC} 
shows the sub-cycle phase-resolved velocity, $\dot{y}_s(t)$, together with the corresponding driving force, $f_y(t)$, over a period $0 \leq t \leq T$ for the flapping frequency $1/T=1\,\text{Hz}$ corresponding to the peak average locomotion velocity. 
At the start of each stroke, $t/T=\{0,0.5\}$, the swimmer initially continues to move in the direction set by the preceding stroke, despite the reversal of the motion of the wings. During this period, the work done by the driving force is negative ($\dot{y}_sf_y<0$), which decelerates the swimmer over a time interval $\Delta t/T\approx 0.1$, until $\left|\dot{y}_s\right|=0$. This behaviour suggests that the swimmer’s inertia cannot be neglected. If inertia were negligible, the swimmer would instantly follow the direction of wing motion, $\dot{y}_s$, and its velocity would change sign immediately at $t/T = \{0, 0.5\}$. 

The swimming motion observed in our experiment occurs when the space-fixed swimmer moves relative to the frictionless container filled with hydrogel granulate. The effective mass associated with this motion is, therefore, that of the container and the hydrogel granulate, rather than the mass of the swimmer itself. Consequently, our experimental setup corresponds to a free swimmer whose density exceeds that of the ambient fluid by orders of magnitude. The high density should only affect the swimmer's inertia, not its buoyancy. Therefore, a free-swimmer experiment that corresponds to our fixed-swimmer setup should be carried out in the absence of gravity. 

A closely related system was investigated by \citet{gonzalez2009reciprocal,hubert2021scallop} who found that reciprocal swimming is possible in a Newtonian fluid within the Stokes regime, provided the swimmer's density is much larger than that of the fluid, $\rho_\text{s}\gg\rho_\text{f}$. In this case, the swimmer-based Reynolds number, $Re_\text{s}=\rho_\text{s}L^2/(T\eta)$, is finite, whereas the Reynolds number of the fluid, $Re_\mathrm{f}=\rho_\mathrm{f}L^2/(T\eta)$, is negligible. Here, $L$ denotes the characteristic length of the swimmer and $\eta$ the viscosity of the fluid. The analytical result  by \citet{gonzalez2009reciprocal} then predicts a cycle-averaged net velocity in the $-y$ direction that increases with the flapping frequency, $1/T$.

Although we acknowledge the substantial differences between the system examined by \citet{gonzalez2009reciprocal} and our own, we note that the direction of motion (along $-y$) and the frequency dependence are similar, at least for low frequencies up to $1/T\approx1\,\text{Hz}$, cf. \autoref{fig:dispB}. At higher frequencies, however, the viscoelastic properties of the hydrogel become significant, and local density variations within the granulate arise due to the rapid motion of the wings. These effects will be discussed in detail in the following sections.



\section{Dynamics of swimming in hydrogels}
\label{sec:dym}

\subsection{Inertia}
\label{sec:inertia}

\begin{figure}[htbp]
\centerline{\includegraphics[width=0.95\linewidth]{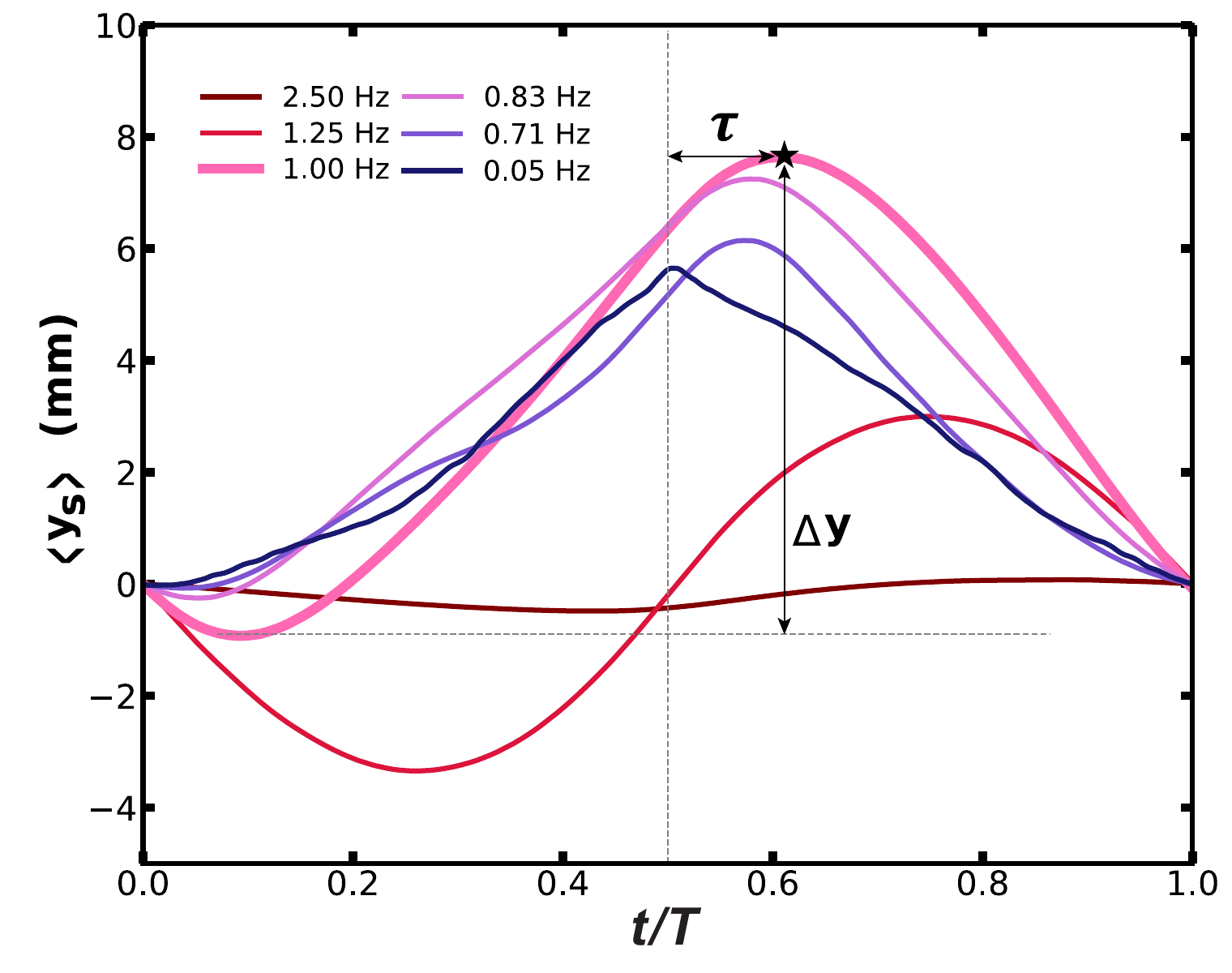}}
\caption{\label{fig:inertiaA} 
Averaged relative displacement of the swimmer as a function of the phase, $t/T$, for various flapping frequencies. The vertical dotted line separates the opening and closing half-cycles. The figure also defines the (positive) peak location, $\tau$, and the peak-to-peak amplitude, $\Delta y$, that depend on the flapping frequency, see \autoref{fig:inertiaB}.}
\end{figure}

The dynamics of the system are determined by the accelerated mass, the flapping period $T$, and the viscoelastic properties of the granulate, which relates the swimmer to the granulate-filled container. In our experiment, the swimmer is fixed in space, therefore, the relevant mass is the mass of the movable granulate-filled container. For a very low frequency, $1/T=0.05\,\text{Hz}$, the container follows the flapping wings approximately, with no noticeable net displacement over one period, as evidenced by a symmetric displacement profile for the opening and closing half-periods, see \autoref{fig:inertiaA}. 
The phase-resolved displacement was obtained by averaging over all periods, disregarding the initial transient. Note that the net transport per cycle, shown in \autoref{fig:dispA}, is much smaller than the typical peak displacement. With its accelerated mass and the restoring force from the granulate, the container may thus be regarded as a driven nonlinear oscillator, oscillating about the swimmer's position with a superimposed slow drift. For slow flapping, $1/T=0.05\,\text{Hz}$, the force acting on the container is small, and the container follows the driving flaps without appreciable deformation of the granular particles. Consequently, the trajectory $\left< y_s \right>(t/T)$ is symmetric between the two half-cycles. For fast flapping ($1/T=2.5\,\text{Hz}$), the container cannot follow the driving motion owing to its inertia. The wings strongly deform the granular particles, but their low elastic modulus prevents sufficient force transfer to accelerate the container appreciably, thus, $\Delta y\to 0$. 

The inertia-based mechanism described above introduces a phase shift $\tau$ between the wing actuation and the relative displacement of the swimmer and container. While the phase shift $\tau$ increases with flapping frequency, the amplitude of the induced oscillation, $\Delta y$ (defined in \autoref{fig:inertiaA}), varies non-monotonically. In the limit $1/T \to 0$, the container follows the driving motion and $\Delta y$ approaches a constant value, $\Delta y^{(0)}$. In the opposite limit $T \to 0$, $\Delta y \to 0$ owing to the low elastic modulus of the medium. At intermediate frequency, $1/T$, we observe amplitudes $\Delta y(T) > \Delta y^{(0)}$, resembling resonant oscillations. 

\begin{figure}[htbp]
\centerline{\includegraphics[width=0.95\linewidth]{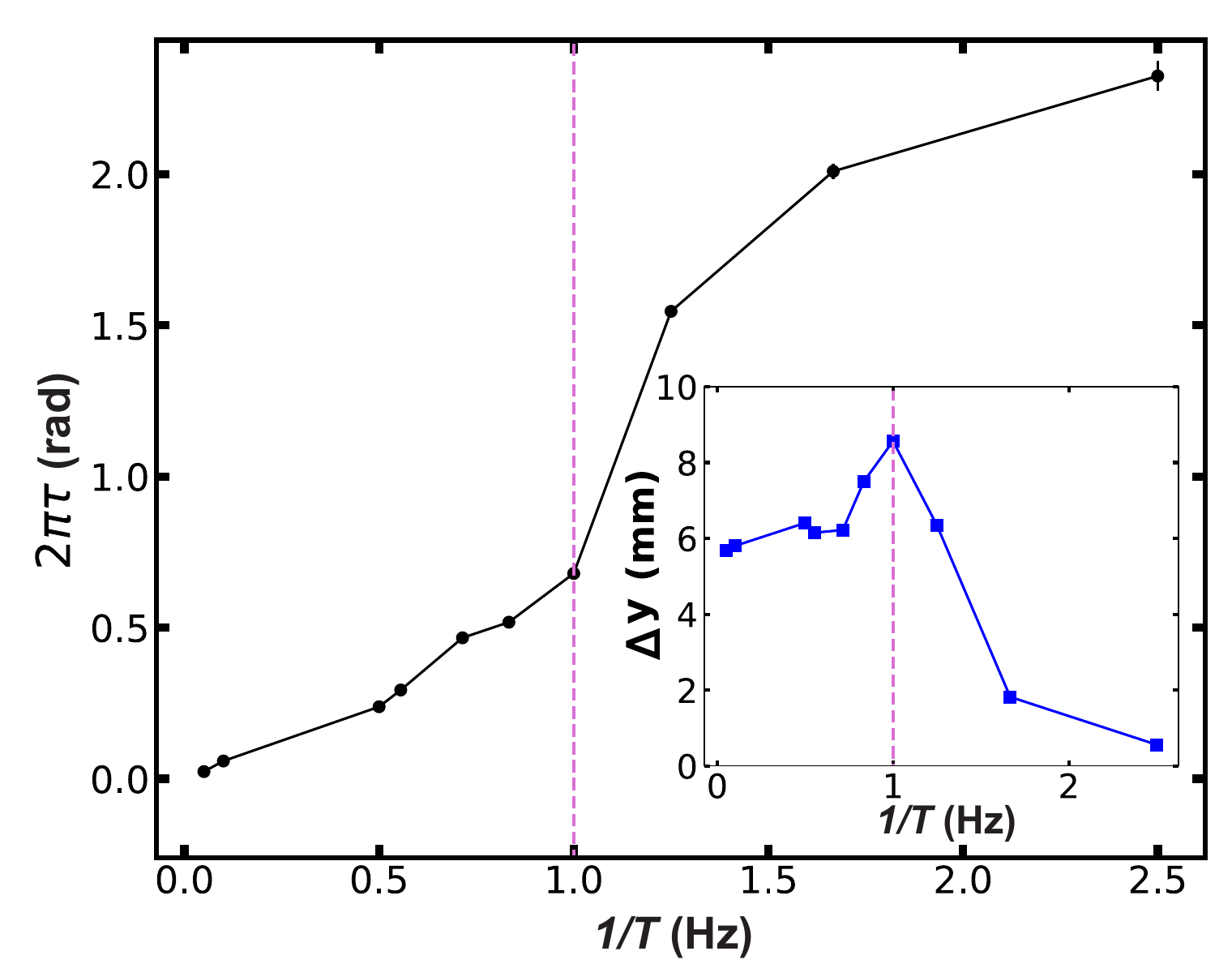}}
\caption{\label{fig:inertiaB} 
Phase shift $\tau$ and peak-to-peak amplitude $\Delta y$ (inset) as functions of the flapping frequency. The vertical dashed lines mark the frequency of maximal locomotion (see \autoref{fig:dispA}). The standard errors in $2\pi\tau$ and $\Delta y$ are indicated by error bars, which are very small.}
\end{figure}

This analysis is supported by \autoref{fig:inertiaB}, which shows the phase shift $2\pi\tau$ and the peak-to-peak amplitude $\Delta y$ as functions of the flapping frequency $1/T$. The dotted lines reveal that the frequency at which $\Delta y$ reaches its maximum coincides with the frequency of fastest locomotion, while the phase shift experiences a rapid increase from $2\pi\tau\approx\pi/4$ to $2\pi\tau\approx\pi/2$. These trends in $\Delta y$ and $\tau$ resemble those of a periodically driven damped oscillator, with resonance occurring near $1\,\text{Hz}$, which is the frequency of maximum locomotion velocity plotted in \autoref{fig:dispB}. 

To clarify this coincidence, we examine the nature of the forces exerted on the swimmer by the hydrogel medium.

\subsection{Viscoelastic relaxation in the hydrogel}
\label{sec:visco}

Depending on the deformation rate, hydrogels exhibit both viscous and elastic behaviours. The hydrogel's response determines the interaction force and, thus, the swimmer's dynamics. We examined the viscoelastic response in relaxation experiments. In case (a), with the air bearings blocked to prevent reservoir displacement, the swimmer’s wings were suddenly opened from $\theta=20^\circ$ to $\theta=80^\circ$ at $t=0$, and the subsequent relaxation of the force (magnitude) on the swimmer, $F_y$, was recorded. Case (b) was analogous, except that the wings were suddenly closed from $\theta=80^\circ$ to $\theta=20^\circ$ at $t=0$. 

\begin{figure}[htbp]
\centerline{\includegraphics[width=0.95\linewidth]{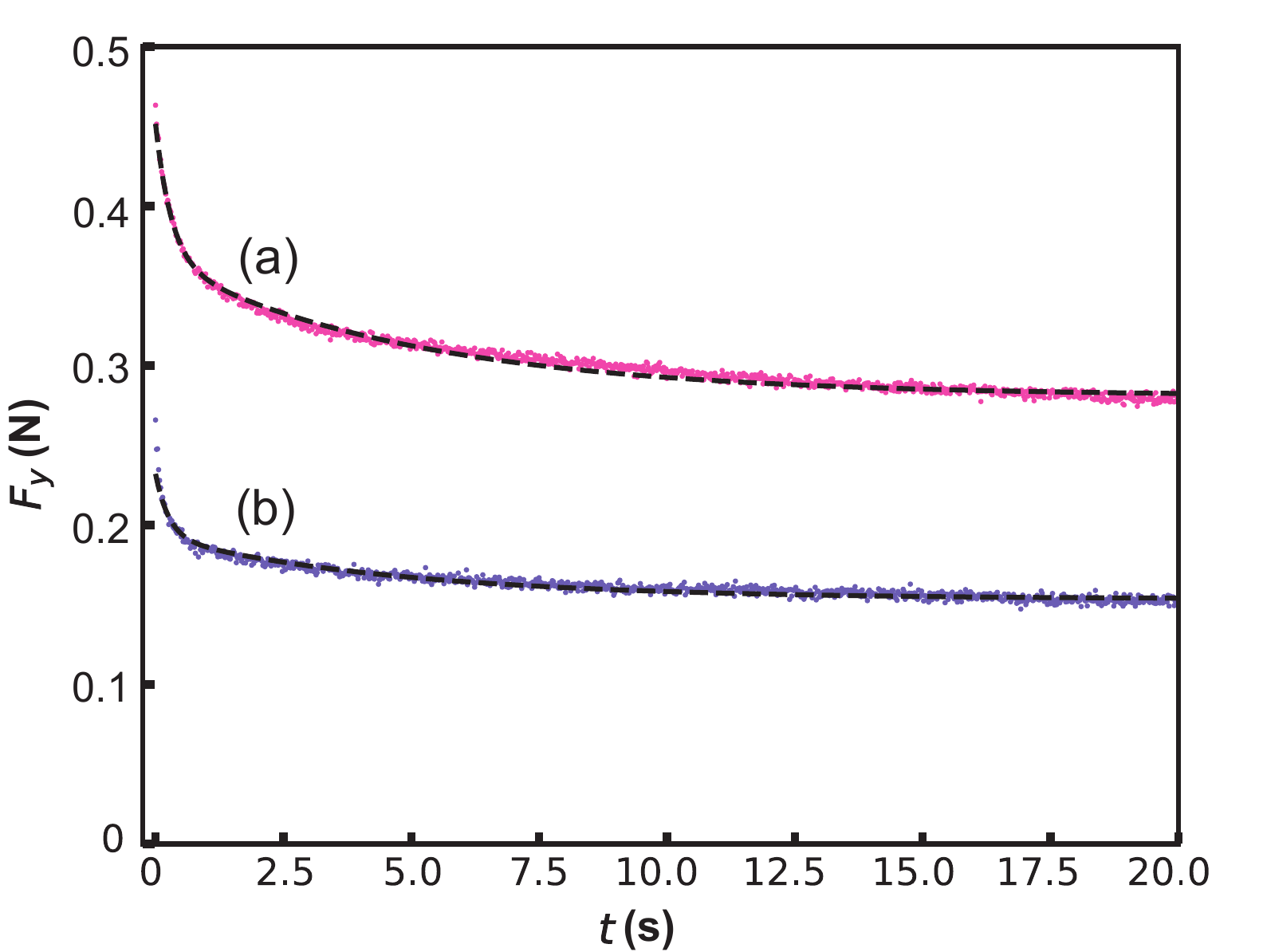}}
\caption{\label{fig:relx} Relaxation of the swimming force following a sudden stroke. For opening (a), the data is fitted to \autoref{eq:relax} with $\alpha=0.085\,\text{N}$ $\tau_1=0.31\,\text{s}$, $\tau_2=5.02\,\text{s}$, and $F_\infty=0.281\,\text{N}$. For closing (b), we obtain $\alpha=0.039\,\text{N}$, $\tau_1=0.26\,\text{s}$, $\tau_2=4.72\,\text{s}$, and $F_\infty=0.154\,\text{N}$.
}
\end{figure}

\autoref{fig:relx} shows the measured forces, 
which exhibit a rapid decay within the first few seconds followed by a slower relaxation toward a long-term residual value, reflecting the combined viscous and elastic response of the medium. 

We fitted the function  
\begin{equation}
F_y = \alpha \left(e^{-t/\tau_1}+e^{-t/\tau_2}\right) + F_\infty\,,
\label{eq:relax}
\end{equation}
to the measured forces over $t\in[0, 20]\text{s}$ to obtain the relaxation times: $\tau_1$ associated with the viscoelastic response of the hydrogel material, and $\tau_2$, associated with particle rearrangements, with $\tau_1 < \tau_2$. Averaging over opening and closing strokes gives $\tau_1=0.29\,\text{s}$ and $\tau_2=4.87\,\text{s}$. 
Remarkably, the peak in backward locomotion at $1\,\text{Hz}$ (see \autoref{fig:dispA}) lies between $1/\tau_1 = 3.4\,\text{Hz}$ and $1/\tau_2 = 0.21\,\text{Hz}$, indicating that viscous effects contribute to the swimmer’s motion. The residual force $F_\infty$ also contributes to  the initial force. On the experimental time scale, $T<20\,\text{s}$, it is effectively elastic.

\subsection{Variation of density in hydrogel media}
\label{sec:xray}
\begin{figure}[htbp]
\centerline{\includegraphics[width=1.0\linewidth]{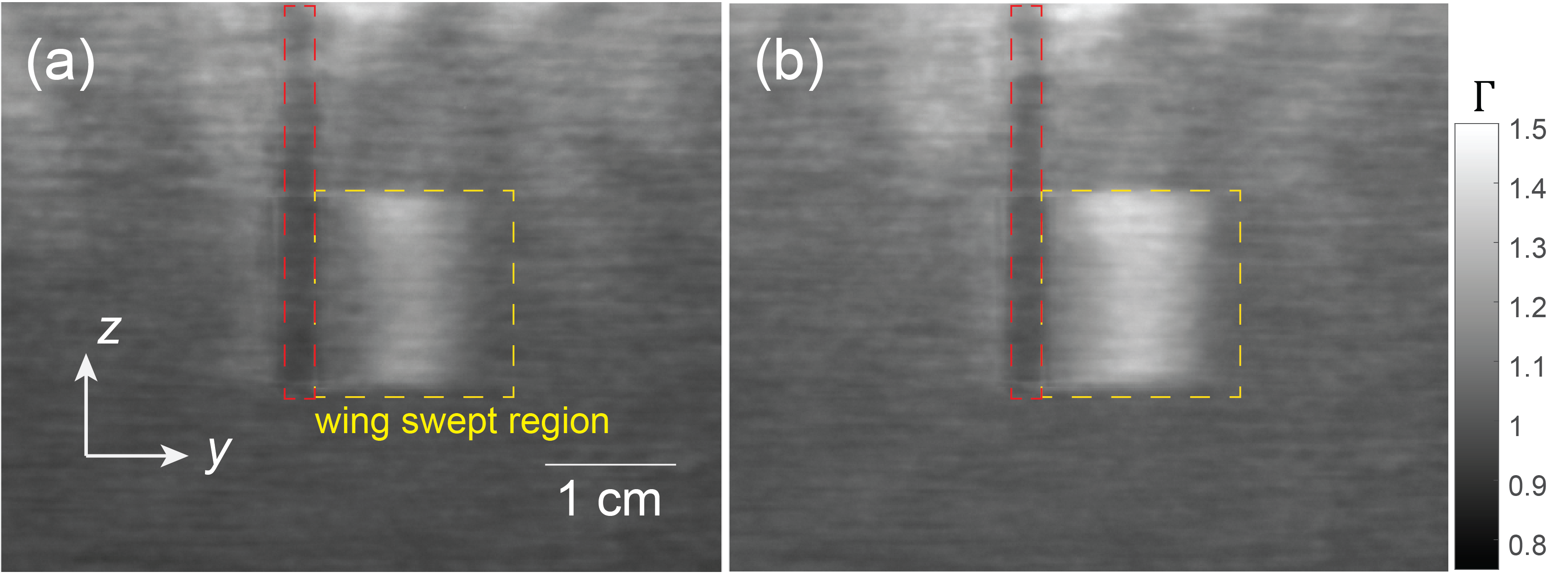}}
\caption{\label{fig:xray} Time-averaged X-ray radiograms of the swimmer flapping with frequency $1/T=0.05\,\text{Hz}$ (a) and $1/T=0.5\,\text{Hz}$ (b). Yellow boxes mark the regions swept by the wings, and red boxes indicate the position of the mounting rod. A higher intensity $\Gamma$ corresponds to a lower density in the hydrogel medium.
}
\end{figure}

Rearrangement of the hydrogel particles, characterized by the relaxation time $\tau_2$, occurs on a slower time scale than the wing motion, leading to densification of the medium in front of the moving wing and dilatation behind it. Whereas densification is limited by the material density of the hydrogel particles, dilatation is not, since air-filled voids may form between the granular particles. Consequently, the average density of the hydrogel granular medium over a flapping period is smaller in the volume swept by the wings than in regions far from the wings. Particle reordering counteracts this effect. Therefore, the difference in density is more pronounced for faster flapping, evidenced by \autoref{fig:xray},
showing X-ray radiographs of the swimmer and the hydrogel granulate in the $y-z$ plane (see \autoref{fig:exp}). The radiographs ($110\,\text{kV}$, $400\,\mu\text{A}$)  were recorded at $10\,
\text{fps}$ and averaged over all frames acquired during at least five flapping cycles. 

Comparison of the X-ray images in panels (a) and (b) shows that the granulate density in the region swept by the wings (yellow box) is lower during fast flapping ($1/T=0.5\,\text{Hz}$) than during slow flapping ($1/T=0.05\,\text{Hz}$). (Because of speed limitations in the X-ray imaging, radiograms could not be obtained for faster wing rotations.) Thus, owing to the relaxation time $\tau_2$, voids form in the vicinity of the wings due to the viscoelasticity of the hydrogel granulate. For $T \sim \tau_2$, the voids can be at least partially refilled during each stroke by particle rearrangements, leading to hysteresis in the swimming cycle despite the reciprocal wing motion.  For $T \ll \tau_2$, particle reordering cannot keep pace with the wing motion, so voids persist during both opening and closing strokes, and symmetry remains unbroken.

\subsection{Implications for locomotion}
From the examination of inertial and viscoelastic effects in Sections \ref{sec:inertia}, \ref{sec:visco}, and \ref{sec:xray}, a consistent picture emerges for the propagation of a swimmer in a hydrogel granular medium, as shown in \autoref{fig:dispA}. To support the subsequent discussion, we introduce an idealized dynamical equation for the swimmer. Although the precise mathematical form of each term is not guaranteed, the equation should at least capture the essential features of the dynamics:
\begin{equation}
m\frac{d^2y_s}{dt^2}+\gamma(\theta)\frac{dy_s}{dt}+\kappa(\theta)(y_s-y_0(t))=f_\text{wing}(\theta)\,.
\label{eq:dynamic}
\end{equation}
Here $m$ is the mass of the movable container, $\gamma$ is the drag coefficient, $\kappa$ an elastic constant, and $f_\text{wing}$ the driving generated by to the wing motion. All terms depend on the instantaneous wing angle $\theta$. In addition, the coefficients $\gamma$ and $\kappa$, as well as the driving force $f_\text{wing}$, depend on the density of the granular medium in a complex manner that is not addressed here. In the stationary state, the mean position $y_0$ drifts at the rate $\Delta y_s/T$, as plotted in \autoref{fig:dispB}. We now discuss the system in three characteristic flapping regimes:
\begin{itemize}[left=0em]
\item For $1/T\gg1\,\text{Hz}$ viscous effects cease as $T\ll \tau_1 < \tau_2$. Here, $\gamma \approx 0$, so the swimmer effectively flaps in an elastic medium while generating long-lasting voids. Owing to the absence of relaxation, $y_0(t) \approx 0$, and \autoref{eq:dynamic} therefore describes harmonic oscillations without locomotion.

\item For $1/T\ll 1\,\text{Hz}$, inertia does not influence swimming, as evidenced by the negligible phase shift in \autoref{fig:inertiaB}.
In this regime, the inertial term $m\,d^2y_s/dt^2$ is obsolete, and \autoref{eq:dynamic} depends only on the wing angle $\theta$. At very low frequencies, particle rearrangements in the hydrogel granulate are fast enough to refill the voids during each stroke, making \autoref{eq:dynamic} effectively density independent. This situation resembles the original scallop theorem, which prohibits locomotion despite strong elastic effects, as seen in \autoref{fig:relx}.

\item For $1/T\approx 1\,\text{Hz}$, viscous and elastic effects coexist, and inertia becomes significant. In this regime, all terms in \eqref{eq:dynamic} contribute, with the microscopic origin linked to hysteresis in the granular density caused by partial refilling of voids on the time scale $\tau_2 \sim T$. 
While the interplay between inertia and viscous effects results in the net swimmer displacement per cycle, it is likely that the interplay between elastic and inertial effects amplifies the net displacement magnitude via a behaviour that resembles resonance, as shown in \autoref{fig:inertiaB}. 
This facilitates the energy transfer between the swimmer and the medium due to the phase shift (\autoref{fig:dispC} and \autoref{fig:inertiaB}), allowing $\gamma dy_s/dt$ and $f_\text{wing}$ to act in the same direction after a switch of the wing rotation direction. This eventually results in a maximum net swimmer displacement at $1/T\approx1\,$Hz. 

\end{itemize}

\section{Summary}
\label{sec:sum}
In summary, a scallop-like swimmer with reciprocally flapping wings in a granular bed of hydrogel spheres can achieve backward locomotion within a certain range of flapping frequencies. This behaviour arises from the combined effects of inertia and viscoelasticity of the soft granular hydrogel medium. The locomotion and its underlying mechanisms differ fundamentally from previous observations in idealized rigid granular media, such as plastic beads \citep{Hongyi:2024}.

\begin{acknowledgments}
R. S. and J. W. acknowledge funding by DLR within project 50WK2348. The authors would like to thank Amir Nazemi for fruitful discussions. 
\end{acknowledgments}


\def\bibsection{\section*{\refname}}
\bibliographystyle{jfm}
\bibliography{swim.bib}

\end{document}